\documentclass[twocolumn,preprintnumbers,amsmath,amssymb]{revtex4}

\usepackage{graphicx}
\usepackage{dcolumn}
\usepackage{bm}

\begin{document}

\title{Microstructures and Mechanical Properties of Dense
Particle Gels:\\ Microstructural Characterization}

\author{Iwan Schenker}
 \email{iwan.schenker@alumni.ethz.ch}
\author{Frank T. Filser}%
\author{Ludwig J. Gauckler}%
\affiliation{ Nonmetallic Materials, Department of Materials, ETH
Zurich, Zurich CH-8093, Switzerland
\homepage{http://www.ceramics.ethz.ch}
}%

\author{Tomaso Aste}
\affiliation{Department of Applied Mathematics, RSPhysSE, The
Australian National University, 0200 Australia}%

\begin{abstract}

The macroscopic mechanical properties of densely packed coagulated
colloidal particle gels strongly depend on the local arrangement
of the powder particles on length scales of a few particle
diameters. Heterogeneous microstructures exhibit up to one order
of magnitude higher elastic properties and yield strengths than
their homogeneous counterparts. The microstructures of these gels
are analyzed by the straight path method quantifying quasi-linear
particle arrangements of particles. They show similar
characteristics than force chains bearing the mechanical load in
granular material. Applying this concept to gels revealed that
heterogeneous colloidal microstructures show a significantly
higher straight paths density and exhibit longer straight paths
than their homogeneous counterparts.

\end{abstract}

\maketitle

\section{\label{sec:Intro}Introduction}

Mechanical properties of coagulated colloids are important in many
technical areas. Sediments~\cite{Yun_2007}, ceramic pastes and
suspensions~\cite{Feng_2000}, pharmaceutical formulations such as
cr\`{e}mes and emulsions~\cite{Adeyeye_2002} and some
food~\cite{Marti_2005} are examples, for which it is desirable to
control the mechanical properties.

Recently, an internal gelation method (DCC = direct coagulation
casting)~\cite{Gauckler_1999,Tervoort_2004} was developed to
process electrostatically stabilized colloidal suspensions to
coagulated particle gels. Thereby it was found that this method
permits to control the gels' microstructures for volume fractions
ranging from 0.2 to 0.6. The method allows for an \it in situ\rm,
i.e., undisturbed destabilization of the colloidal suspension by
either changing the pH of the solution ($\Delta$pH-method)
resulting in a ``homogeneous'' microstructure or by increasing the
ionic strength ($\Delta$I-method) leading to a ``heterogeneous''
microstructure~\cite{Wyss_2004}.

\begin{figure}
\includegraphics[width=\linewidth]{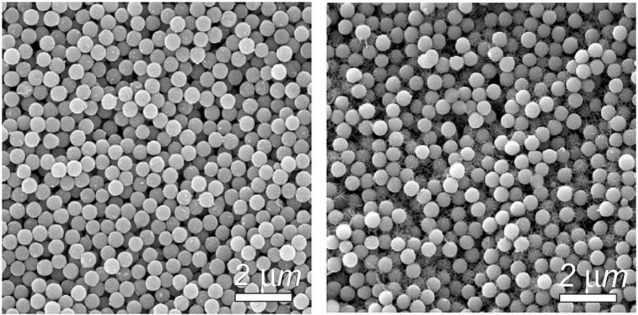}
\caption{Microstructure of coagulated silica suspensions at a
volume fraction of 0.4 formed by the $\Delta$pH- (left) and the
$\Delta$I-method (right) using the DCC process (particle diameter
0.525~$\mu$m). Micrographs obtained by cryogenic scanning electron
microscopy~\cite{Wyss_2004}.} \label{fig:load_Fig1}
\end{figure}

Figure~\ref{fig:load_Fig1} shows cryogenic scanning electron
microscopy (cryo-SEM) pictures of both microstructures. The
particles in the $\Delta$pH-coagulated gel present a highly
homogeneous microstructure whereas in the $\Delta$I-system
inhomogeneities on the length scale of a few particle diameters
are observed. This qualitative observation was quantified by the
three-dimensional pair-correlation function calculated from stereo
cryo-SEM images~\cite{Wyss_2004} and by \it in situ \rm performed
diffusing wave spectroscopy (DWS) experiments during
destabilization~\cite{Wyss_2001}. The structural differences
correspond to differences in the heterogeneity: the first peak of
the pair-correlation function at $r/d = 1$ ($r$ being the distance
between the particles and $d$ the particle diameter) is higher for
the $\Delta$I-system indicating locally denser regions with higher
average coordination number. Also, local maxima are present in the
pair-correlation function for the $\Delta$I-system at $r/d \approx
1.4$ and at $r/d \approx 1.6$ corresponding to characteristic
peaks in hexagonally packed particle arrangements. During the
$\Delta$I-destabilization, particle rearrangements lead to larger
pores and thus to less homogeneous
microstructures~\cite{Wyss_2001}, whereas during the
$\Delta$pH-destabilization, the initially stabilized, liquid-like
microstructure, is ``frozen'', which results in a more homogeneous
microstructure.

Rheological and uniaxial compression experiments on coagulated
colloidal structures obtained by
DCC~\cite{Balzer_1999,Wyss_1_2005} showed that colloids with
heterogeneous microstructures have significantly higher elastic
moduli and yield strengths than their homogeneous counterparts.
Measuring the dynamics of the colloids during destabilization by
DWS confirmed these findings quantitatively using a model for the
storage modulus proposed by Krall and
Weitz~\cite{Wyss_2001,Krall_1998}.

Alkali-swellable polymer (ASP) particles were used in order to
introduce heterogeneities during the pH-destabilization. Small
amounts of 80~nm ASP particles were admixed to the powders under
acidic conditions. These particles swell upon changing pH during
the internal gelling reaction of the DCC process and enfold to
0.7~$\mu$m size producing less homogeneous microstructures. Those
samples with swollen polymer particles showed much higher
mechanical properties in comparison to samples without swellable
polymers and hence more homogenous
microstructures~\cite{Hesselbarth_2000}. In particular, they
present the same high mechanical properties as samples with
heterogeneous microstructures produced by the
$\Delta$I-method~\cite{Hesselbarth_2000}.

\begin{figure}
\includegraphics[width=\linewidth]{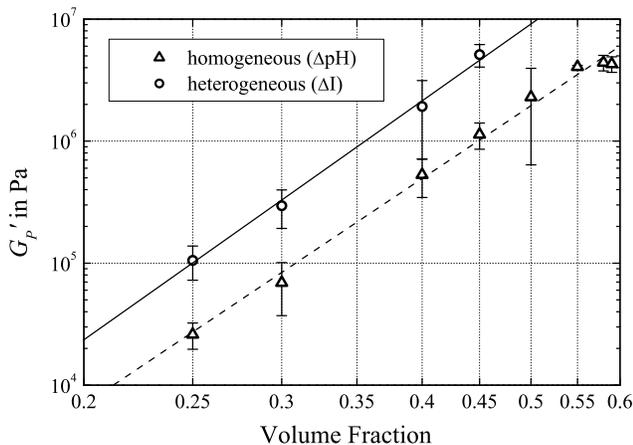}
\caption{Elastic plateau storage modulus $G'_p$ of alumina
particle suspensions (average particle diameter $d_0 =
0.4$~$\mu$m) in dependence of the volume fraction formed by the
$\Delta$pH- and the $\Delta$I-method of the DCC
process~\cite{Wyss_1_2005}.} \label{fig:load_Fig2}
\end{figure}

Examples of the rheologically measured elastic properties of
alumina particle suspensions (average particle diameter $d_0 =
0.4$~$\mu$m) for different volume fractions destabilized by the
$\Delta$pH- and the $\Delta$I-method, respectively, are shown in
Fig.~\ref{fig:load_Fig2}. Almost one order of magnitude higher
elastic plateau storage moduli are measured for heterogeneous
microstructures than for those with homogenous
microstructures~\cite{Wyss_1_2005}.

In summary, strong evidence is given that the differences in
macroscopic mechanical properties of coagulated particle
suspensions are controlled by the differences in microstructure.
An open question is now how these microstructural differences on
particle length scales can have such a dramatic influence on the
mechanical properties. Therefore, a concise quantitative analysis
of microstructures of colloidal particle systems is needed.

In preceding works, various characterization methods, such as the
radial pair-correlation function~\cite{Huetter_2000}, the bond
angle distribution function~\cite{Huetter_2000}, the triangle
distribution function~\cite{Huetter_2000} and the Minkowski
functionals in conjunction with the parallel-body
technique~\cite{Huetter_2003}, were applied to sets of
microstructures generated by Brownian dynamics (BD)
simulations~\cite{Huetter_1999}. These simulations were used to
study the coagulation dynamics and the evolving microstructures in
dense colloidal suspensions and the resulting microstructures
agree well with experiments~\cite{Wyss_2002}.

While the pair-correlation function permits to quantify the amount
of structural rearrangement during the
coagulation~\cite{Huetter_2000}, the analysis using the Minkowski
functionals in conjunction with the parallel-body technique
supplies additional information on the structure's morphology
resolving microstructural differences on a length scale limited by
the largest pore size~\cite{Huetter_2003}. The bond angle
distribution function and the triangle distribution functions are
useful means to examine the local building blocks of the particle
network~\cite{Huetter_2000}. Particular features, as, for example,
peaks in the respective distribution function have successfully
been correlated to the structure's heterogeneity and porosity. The
same conclusion is valid for the Minkowski
functionals~\cite{Huetter_2003}.

All these four methods are good means to compare structures in
terms of their heterogeneity. However, these structural
descriptions do not unambiguously help to understand why more
heterogeneous colloidal structures possess stronger mechanical
properties, as they do not adequately capture the microstructural
characteristics that are responsible for the mechanical properties
of these particle systems.

It is well known from granular matter physics that particulate
systems under mechanical stress carry load via chains of
contacting particles, termed as force chains, as observed in
experiments on granular materials~\cite{Drescher_1972} and in
simulations~\cite{Cundall_1979}. Granular materials, as, for
example, sand piles, that contain large amounts of particles
arranged in chains of contacting particles are expected to possess
higher mechanical properties than those in which the particles
have to rearrange upon applied external load in order to form such
chains. Hence it is plausible that these chains of contacting
particles control the mechanical properties.

The aim of this paper is to adequately analyze the microstructure
of dense colloids and to link microstructural differences to
differences in macroscopic mechanical properties upon applied
mechanical stress. Thereby, the focus is on the identification of
densely packed regions and chains of contacting particles in these
particle networks that may correlate differences in mechanical
properties with differences in microstructures. Our approach is to
analyze the distribution of densely packed regions in each
microstructure using the common neighbor distribution and the
dihedral angle distribution function. Additionally, a new method
called the straight path method, characterizing the quasi-linear
arrangement of particle chains, is introduced. The evaluation of
these characterization methods is performed with regard to their
ability to distinguish quantitatively between homogeneous and
heterogeneous microstructures obtained from
BD-simulations~\cite{Huetter_1999} mimicking well those found in
experiments~\cite{Wyss_2002}.

\section{\label{sec:MatAndMeth}Materials and Methods}

\subsection{Structure Generation}

\begin{figure}
  \includegraphics[width=\linewidth]{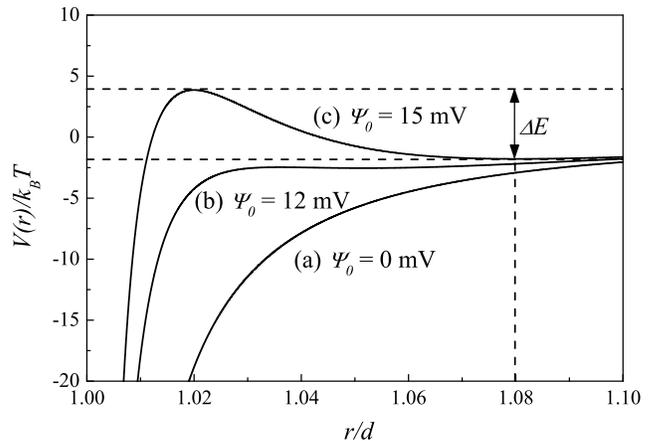}
\caption{DLVO interaction for different surface potentials: (a)
pure van der Waals attraction ($\Delta$pH-destabilization), (b)
$\Delta E = 0 k_B T$ and (c) $\Delta E = 5.65 k_B T$
($\Delta$I-destabilization).} \label{fig:load_Fig3}
\end{figure}

The homogeneous and the heterogeneous microstructures are fully
coagulated colloidal suspensions obtained from Brownian dynamics
(BD) simulations~\cite{Huetter_1999}. The
Derjaguin-Landau-Verweg-Overbeek (DLVO) theory~\cite{Russel_1989}
was used to describe the particle-particle interaction, where
pair-wise particle potential interactions are assumed given by the
sum of the van der Waals attraction $V^{vdw}$
(Eq.~(\ref{eq:load_Vvdw})) and the electrostatic double layer
repulsion $V^{el}$ (Eq.~(\ref{eq:load_Vel})). Thus, $V^{dlvo} =
V^{vdw} + V^{el}$ with

 \begin{equation}
    V^{vdw}(r) = - \frac{A_H}{12}\left[\frac{d^2}{r^2-d^2} + \frac{d^2}{r^2} + 2 \ln \left(\frac{r^2-d^2}{r^2}\right)\right]
 \label{eq:load_Vvdw}
 \end{equation}

\noindent and

 \begin{equation}
    V^{el}(r) = \pi \epsilon_r \epsilon_{\mathrm{0}}
    \left[ \frac{4 k_b T}{z_V e} \tanh \left( \frac{z_V e}{4 k_b T} \Psi_{\mathrm{0}} \right)
    \right]^2
    d \exp \left( -\kappa \{ r-d\} \right),
 \label{eq:load_Vel}
 \end{equation}

\noindent respectively. The DLVO parameters are summarized in
Table~\ref{tab:load_BDparameter} and the potential curves for
various values of the surface potential $\Psi_{\mathrm{0}}$ are
shown in Fig.~\ref{fig:load_Fig3}.

\begin{table}
\caption{Potential Parameters for the Coagulating Suspensions}
\label{tab:load_BDparameter}
\begin{center}
\begin{tabular}{lll}
\hline\noalign{\smallskip}
Parameter & Symbol & Value  \\
\noalign{\smallskip}\hline\noalign{\smallskip}
Hamaker constant of Al$_2$O$_3$ in water  & $A_H$ & $4.76 \times 10^{-20}$ J\\
Particle diameter  & $d$ & 0.5~$\mu$m \\
Relative dielectric constant of water & $\epsilon_{r}$ & 81 \\
Absolute temperature  & $T$ & 293 K \\
Valency of ions  & $z_V$ & 1 \\
Inverse Debye screening length & $\kappa$ & 10$^{8}$ m$^{-1}$ \\
\noalign{\smallskip}\hline
\end{tabular}
\end{center}
\end{table}

For $\Psi_{\mathrm{0}} = 0$ mV, the electrostatic double layer
repulsion is zero and the inter-particle potential is only given
by the attractive van der Waals potential. For $\Psi_{\mathrm{0}}
= 12$ mV a secondary minimum appears and $\Delta$E denotes the
energy barrier between the local maximum and the secondary
minimum. For $\Psi_{\mathrm{0}} = 15$ mV, a repulsive barrier of
$\Delta$E = 5.65 $k_B T$ exists and the secondary minimum is found
at $r = 1.08 d$. The model further contains the frictional Stokes'
drag force and a random Brownian force caused by the suspending
liquid.

We analyze microstructures generated using $\Psi_{\mathrm{0}} = 0$
mV, which correspond to the $\Delta$pH destabilization method and
$\Psi_{\mathrm{0}} = 15$ mV, corresponding to the $\Delta$I
destabilization method. In the following, the
$\Delta$pH-microstructures are labeled as homogeneous and the
$\Delta$I-microstructures as heterogeneous. All structures have
the same volume fraction of 0.4 and contain 8000 spherical
particles forming one percolating cluster. The particles have a
diameter of 0.5~$\mu$m.

\subsection{Structure Characterization Methods}

\subsubsection{Common Neighbor and Dihedral Angle Distribution
Function}

The common neighbor analysis~\cite{Clarke_1993} considers pairs of
particles, referred to as configurations, and determines the
number of particles that are in contact with both particles of the
configuration. In two dimensions, configurations can have at most
two common neighbors. In three dimensions, configurations can have
up to five common neighbors. The number of configurations with n
common neighbors is denoted $CN_n$ with $n=1,\ldots,5$. In this
study, absolute numbers of configurations are compared as the
microstructures analyzed have identical volume fraction, the same
number of particles and equal particle diameters.

\begin{figure}
  \includegraphics[width=\linewidth]{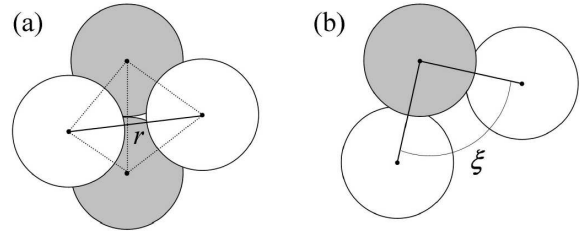}
\caption{Configuration (grey particles) with two common neighbors
(white particles) separated by a distance $r$ and having a
dihedral angle $\xi$. Perspective view (a) and top view (b).}
\label{fig:load_Fig4}
\end{figure}

The common neighbor distribution describes the short range
arrangement of particles. In particular, $CN_0$ is the number of
pairs of contacting particles with no common neighbor. $CN_1$
counts the number of configurations with exactly one neighbor
common to both particles, forming equilateral triangles. For $n$ =
2 the four particles of the configuration form a regular
tetrahedron if the two common neighbors are in contact, and a
generalized tetrahedron, having one longer edge, otherwise.

Configurations with two or more common neighbors ($CN_2,\ldots,
CN_5$) can be further characterized by the dihedral angle
distribution. This distribution analyses the arrangement of the
neighboring particles around the two particles of the
configuration, which form regular triangles with each of their
common neighbors. For $n \geq 2$ there are at least two common
neighbors. In this case, the dihedral angle $\xi$ is defined as
the angle between two planes spanned by triangles belonging to the
same configuration as is shown in Fig.~\ref{fig:load_Fig4}. $\xi$
is measured between one triangle and the triangles formed with
each other common neighbor of the original couple of particles.
Chemistry uses this method to characterize molecular
structures~\cite{Wielopolski_1986}. In the field of granular
matter, the dihedral angle distribution has been used to study the
structural organization and correlations in very large packings of
monodispersed spherical particles~\cite{Aste_2005,Aste_2006}.

For two common neighbors in contact with each other, the four
particles form a regular tetrahedron and the dihedral angle is
70.5$^{\circ}$. This angle constitutes the lower limit of the
dihedral angle distribution. The upper limit is 180$^{\circ}$ when
the four particles are in a plane and form a square.

Formally, the dihedral angle $\xi$ between two common neighbors of
a configuration having a separation distance $r$ between each
other is defined by $\xi = 2 \arcsin \left( \frac{r}{\sqrt{3} d}
\right)$. The dihedral angle distribution is given by $p(\xi +
\delta \xi / 2) = \delta N_{\xi} (\xi, \xi + \delta \xi)$ where
$\delta N_{\xi} (\xi, \xi + \delta \xi)$ is the number of
generalized (open) tetrahedra with a dihedral angle between $\xi$
and $\xi + \delta \xi$~\cite{Aste_2006}.

\subsubsection{Straight Path Method}

\begin{figure}
  \includegraphics[width=\linewidth]{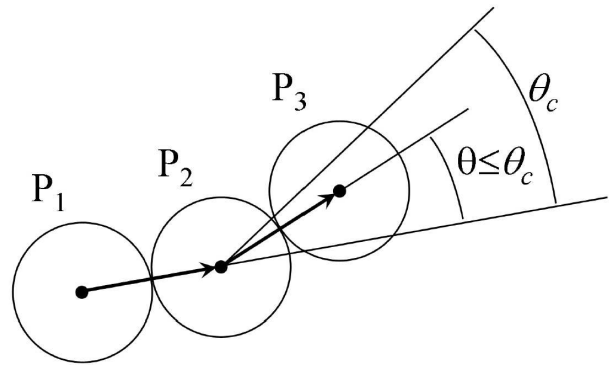}
\caption{Chain of three particles forming a straight path.}
\label{fig:load_Fig5}
\end{figure}

A straight path is a quasi-linear chain of contacting particles.
The shortest chain consists of three particles. In order for three
particles to form a straight path they have to fulfill the
following condition: The angle $\theta$ between the vector
connecting the first two particles (P$_1$, P$_2$) and the vector
connecting the second two particles (P$_2$, P$_3$) must be smaller
than a chosen threshold angle $\theta_c$. This is exemplified in
Fig.~\ref{fig:load_Fig5}.

Analogously, a straight path of length $l$ is one of length $l-1$
that has another particle attached to one end and whose connection
vector fulfills the condition to stay in a direction within a cone
of angle $\theta_c$ from the vector connecting the previous two
particles. The distribution of straight path lengths is obtained
by determining the number of straight paths $SP_l$ of length
$l=3,\ldots,l_{max}$, with $l_{max}$ the longest path under
consideration. Absolute numbers are calculated and compared in the
distribution of $SP_l$.

The angle $\theta_c$ determines the quasi-linearity and specifies
the maximally allowed deflection from an absolute linear
arrangement of the particles. The dependence of the straight path
statistic on the angle $\theta_c$ is presented as part of the
results section and is used in order to choose a suitable angle
$\theta_c$. The distribution of straight paths for the homogeneous
and heterogeneous microstructures is compared. The average path
length $SP_{mean} = \Sigma_{l\geq 3} l SP_l / \Sigma_{l\geq 3}
SP_l$ and the number of paths longer than a chosen length $l_0$ is
determined. The latter is defined by $SP_{l\geq l_0} =
\Sigma_{l\geq l_0} SP_l$ ($l_0$ = 4 and 5) and permits to quantify
differences in the number of straight paths with longer lengths,
neglecting paths shorter than $l_0$ particles.

\section{Results and Discussion}
\label{ResAndDisc}

\subsection{Common Neighbor Analysis}

\begin{figure}
  \includegraphics[width=\linewidth]{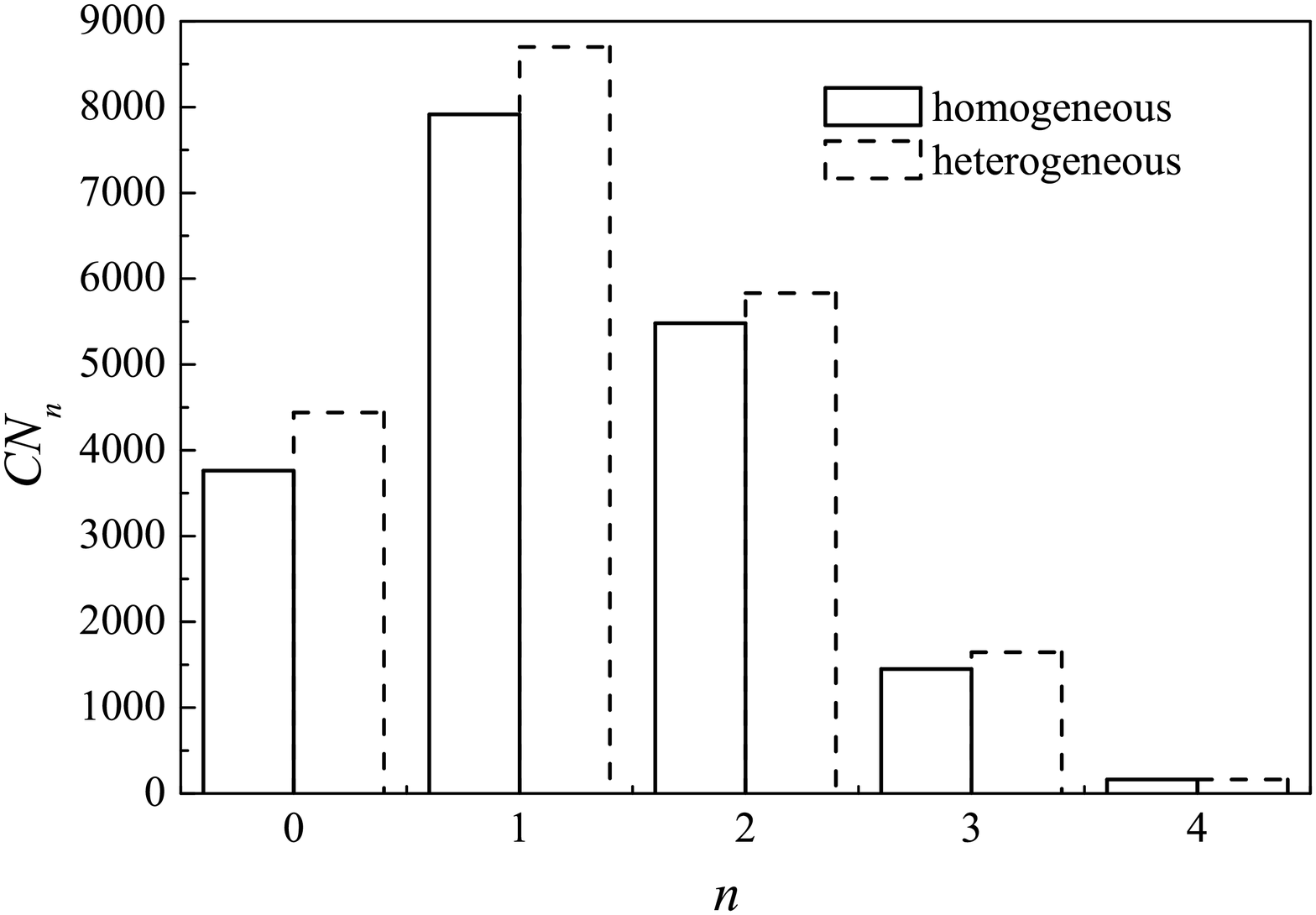}
\caption{Common neighbor distribution for the homogeneous and
heterogeneous microstructures: number of configurations $CN_n$ in
dependence of the number of common neighbors $n$.}
\label{fig:load_Fig6}
\end{figure}

The number of particle configurations $CN_n$ is higher for the
heterogeneous than for the homogeneous microstructure for
$n=0,\ldots,3$ (Fig.~\ref{fig:load_Fig6}). This can easily be
understood as the mean coordination number of a particle in the
heterogeneous microstructure is with 5.2 roughly 10\% larger than
in the homogeneous microstructures with 4.7. When normalized by
the mean coordination number, the distributions collapse,
indicating that the relative distribution of common neighbor
configurations is identical for both microstructures. Thus, the
common neighbor distribution does not capture in a suitable way
the structural differences between these microstructures.

\subsection{Dihedral Angle Distribution}

\begin{figure}
  \includegraphics[width=\linewidth]{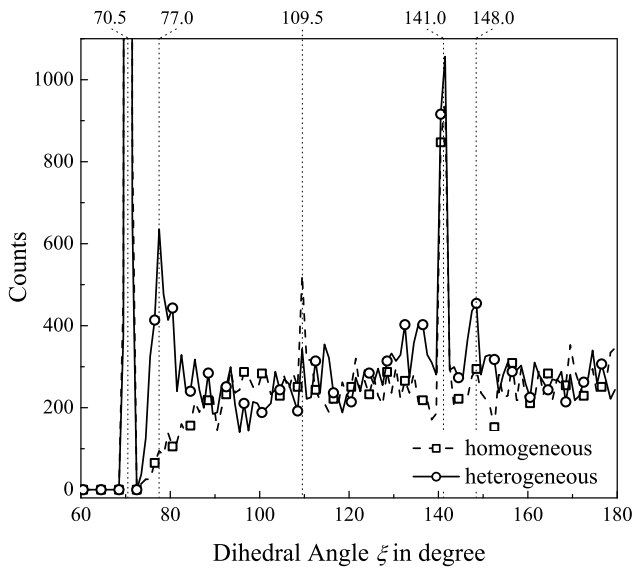}
\caption{Dihedral angle distribution for the homogeneous and
heterogeneous microstructures. For clarity purposes, only every
fourth data point is indicated by a symbol.} \label{fig:load_Fig7}
\end{figure}

The dihedral angle distributions of both microstructures show a
characteristic peak at 70.5$^{\circ}$ (Fig.~\ref{fig:load_Fig7})
corresponding to configurations that form regular tetrahedra with
two common neighbors. The peak height is the same for both
microstructures. The heterogeneous microstructure has a peak at
77$^{\circ}$, which is missing for the homogeneous one. This angle
corresponds to a separation distance between the common neighbors
of $1.08 d$ which is the distance between particles trapped in the
secondary minimum (Fig.~\ref{fig:load_Fig3}).

At $\xi = 109.5^{\circ}$ the homogeneous microstructure has a peak
that corresponds to a common neighbor separation of $1.41 d
\approx \sqrt{2}d$, which is absent in the case of the
heterogeneous microstructure. This separation distance corresponds
to two configurations forming a square and sharing two common
neighbors. An additional characteristic peak is found at
141$^{\circ}$ having roughly the same height for both
microstructures and corresponding to two juxtaposed tetrahedra.
The peak at 148$^{\circ}$ is solely present for the heterogeneous
microstructure and corresponds to one regular tetrahedron
juxtaposed to one with a distance of $1.08 d$ between the common
neighbors.

In conclusion, even if there are sizable differences between
homogeneous and heterogeneous structures, the absolute number of
regular tetrahedra is virtually equal for both microstructures and
the analysis does not reveal any structural difference that could
explain the better mechanical properties of the heterogeneous
microstructures. Thus, the common neighbor and the dihedral angle
distributions are not suited to link the microstructural
differences to the different mechanical properties.

\subsection{Straight Path Analysis}

\begin{figure}
  \includegraphics[width=\linewidth]{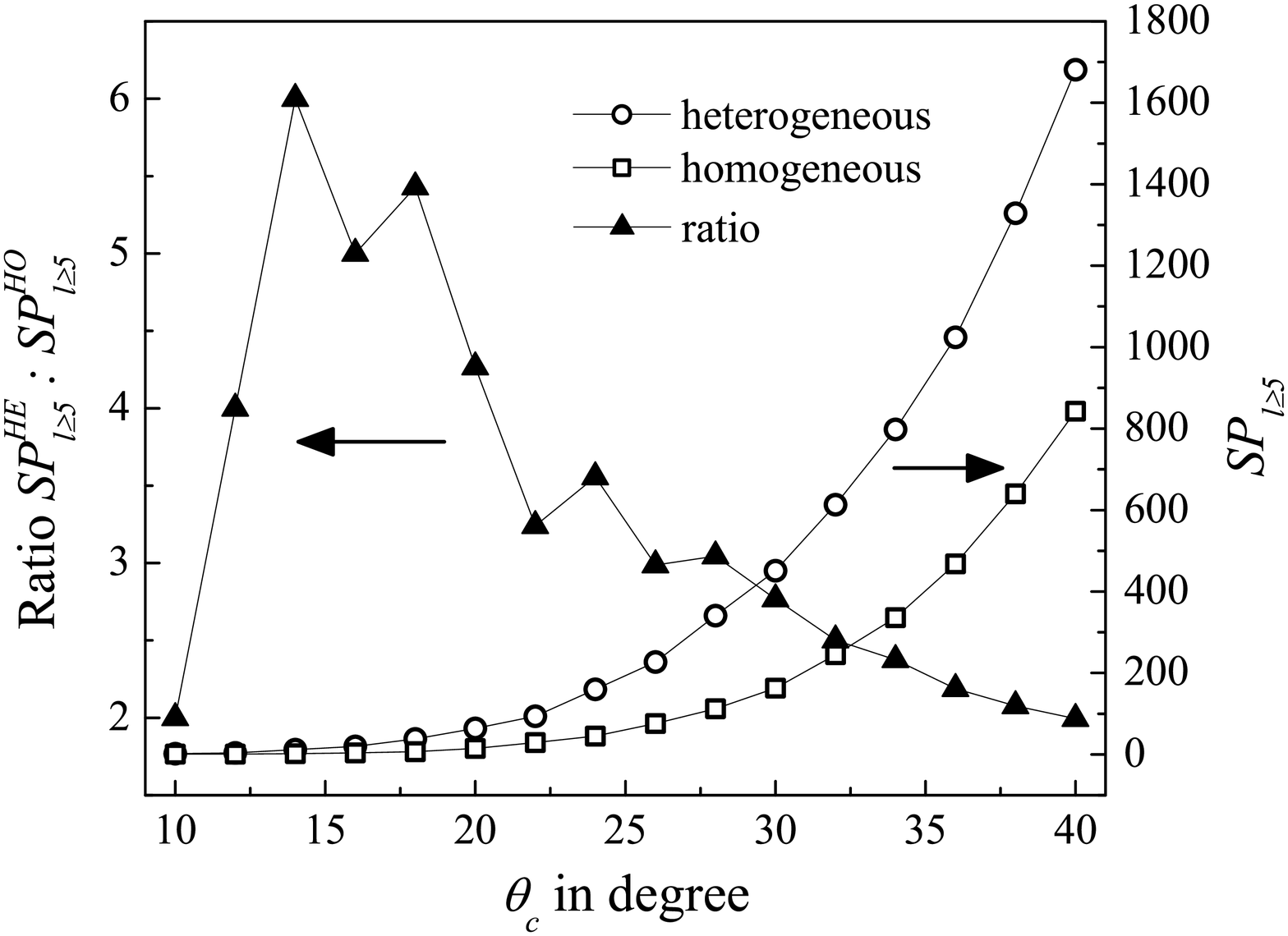}
\caption{Number of straight paths for both microstructures longer
or equal to five ($SP_{l=5}^{HE}$ and $SP_{l=5}^{HO}$, open
symbols) and their ratio ($SP_{l=5}^{HE} : SP_{l=5}^{HO}$,
triangles) as a function of the threshold angle $\theta_c$.}
\label{fig:load_Fig8}
\end{figure}

The straight path analysis depends on the threshold angle
$\theta_c$, hence, a suitable value $\theta_c$ has to be
determined. For small angles $\theta_c$ only a small number of
straight paths are expected because only a few particle chains
satisfy the criterion to be a straight path. For large angles
$\theta_c$ many straight paths are found but the physical meaning
of a straight path is lost as the path becomes less ``straight''
and thus loses its load bearing capacity. Also for increasing
$\theta_c$ the ratio of the number of paths with equal lengths
between the heterogeneous and the homogeneous microstructure
decreases. Figure~\ref{fig:load_Fig8} shows the number of straight
path longer or equal to five particles for the homogeneous
($SP_{l\geq 5}^{HO}$) and for the heterogeneous microstructure
($SP_{l\geq 5}^{HE}$) as well as their ratio ($SP_{l\geq 5}^{HE} :
SP_{l\geq 5}^{HO}$) in dependence of the cone angle $\theta_c$
ranging from 10$^{\circ}$ to 40$^{\circ}$. The lines serve as
guide for the eyes.

For $\theta_c$ between 14$^{\circ}$ and 18$^{\circ}$ the ratio
$SP_{l\geq 5}^{HE} : SP_{l\geq 5}^{HO}$ shows a maximum. However,
below 22$^{\circ}$ the number of straight paths is quite small
($SP_{l\geq 5}<100$) for both microstructures. Toward larger
angles the absolute number of straight paths is increasing while
the ratio $SP_{l\geq 5}^{HE} : SP_{l\geq l5}^{HO}$ is decreasing.
The angle $\theta_c = 30^{\circ}$ was chosen as this angle gives
reasonable values for both the ratio ($SP_{l\geq 5}^{HE} :
SP_{l\geq 5}^{HO} \approx 3$) and absolute number of paths
($SP_{l\geq 5}^{HE}= 451$ and $SP_{l\geq 5}^{HO}= 163$).

\begin{figure}
  \includegraphics[width=\linewidth]{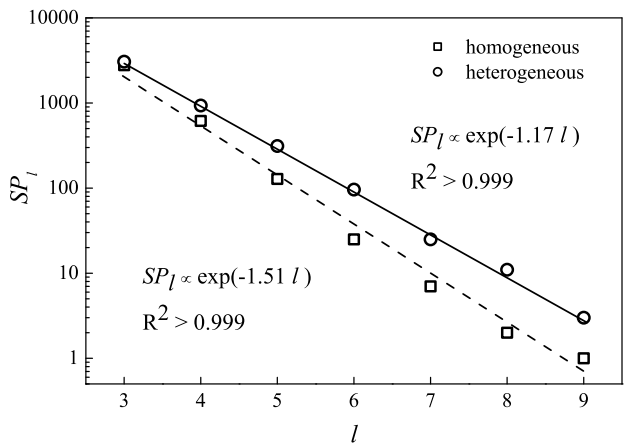}
\caption{Number of straight paths $SP_l$ in dependence of the
straight path length $l_{SP}$ for the homogeneous and
heterogeneous microstructures (squares and circles, respectively)
and exponential fits (dashed and solid line, respectively).}
\label{fig:load_Fig9}
\end{figure}

For both microstructures the distribution of straight paths shows
an exponential decrease with straight path length $l_{SP}$
(Fig.~\ref{fig:load_Fig9}). Both microstructures have
approximately the same number of straight paths of length three.
Toward longer path lengths the microstructures present a diverging
behavior: the heterogeneous microstructure has more paths of
longer length than the homogeneous microstructure. The
distributions were fitted by an exponential law using $SP_l
\propto \exp (-\lambda l)$, with $\lambda$ the characteristic
inverse path length. A remarkably good fit of the data is obtained
for both microstructures indicated by the values close to one for
the correlation coefficients R$^2$. The higher exponent $\lambda$
found for the homogeneous microstructure demonstrates the more
rapid decrease of the number of straight paths toward longer path
lengths than in the case of the heterogeneous microstructure.

The values for the average path lengths ($SP_{mean}$) and the
numbers of paths longer or equal to four ($SP_{l\geq 4}$) and five
($SP_{l\geq 5}$) are summarized in Table~\ref{tab:load_SP}. The
average path length for the heterogeneous microstructure is only
5\% longer than for the homogeneous microstructure because
$SP_{l=3}$ dominates the statistics and is almost identical for
both structures. On the other hand, in the heterogeneous
microstructure longer chains $SP_{l\geq 4}$ and $SP_{l\geq 5}$ are
found 1.8 and 2.8 times more often, respectively, than in the
homogeneous one.

\begin{table}
\caption{Average path length $SP_{mean}$ and numbers of paths
longer or equal to four ($SP_{l\geq 4}$) and five ($SP_{l\geq 5}$)
for both microstructures} \label{tab:load_SP}
\begin{center}
\begin{tabular}{lll}
\hline\noalign{\smallskip}
 & Homogeneous & Heterogeneous \\
\noalign{\smallskip}\hline\noalign{\smallskip}
$SP_{mean}$ & 3.28 & 3.45 \\
$SP_{l\geq 4}$ & 779 & 1385 \\
$SP_{l\geq 5}$ & 163 & 451 \\
\noalign{\smallskip}\hline
\end{tabular}
\end{center}
\end{table}

To conclude, the number of straight paths in the homogeneous and
heterogeneous microstructures follows an exponential distribution
and a higher characteristic inverse path length $\lambda$ is found
for the homogeneous microstructure. The heterogeneous
microstructure contains significantly more straight paths of
longer lengths than the homogeneous one: in absolute numbers twice
as many paths with a length $\geq 4$ and three times as many
having a length $\geq 5$ are observed. These are very significant
differences between the two microstructures.

Our results suggest that the differences in the straight path
distributions are characteristic for the better mechanical
properties of the heterogeneous microstructures. Indeed, in
granular materials load is transmitted via force chains that are
quasi-linear substructures of the particle network. This has been
experimentally verified using photo-elasticity
techniques~\cite{Drescher_1972} and computationally modelled using
the discrete element method~\cite{Cundall_1979}. Also, the
computational study of the force chain network in 2D granular
assemblies of polydispersed particles subjected to indentation by
a rigid flat punch has shown that the force chain length follows
an exponential distribution~\cite{Peters_2005}.

\section{Conclusions}
\label{Conclusions}

Different chemical pathways during the internal destabilization of
colloidal suspensions lead to wet ceramic green bodies with
drastically different mechanical properties. The microstructures
of these destabilized colloids reveal differences in the
heterogeneity of the particle arrangements on a length scale in
the order of a few particle diameters. Various microstructural
characterization methods, i.e., the common neighbor distribution,
the dihedral angle distribution and the straight path method, have
been applied in order to analyze homogeneous and heterogeneous
microstructures generated by Brownian dynamics simulations, which
were shown to agree well with the experimentally determined
microstructures of such coagulated colloidal particle systems.

Toward our goal, to establish a correlation between the
microstructural differences and the differences in macroscopic
mechanical properties, it was found that the common neighbor and
the dihedral angle distributions do not discriminate enough the
differences between the homogeneous and heterogeneous
microstructures and can therefore hardly account for the large
differences in mechanical properties. The newly introduced
straight path method reveals significantly more straight paths of
longer lengths for the heterogeneous than for the homogenous
microstructure: twice as many straight paths of length $\geq 4$
particles and three times as many of length $\geq 5$ particles are
found in the heterogeneous microstructure. These differences in
straight path number and length seem suitable to characterize and
to differentiate between these structures and these differences
are considered to be sufficiently large to account for the
differences in mechanical properties. The straight paths may
capture best the characteristic microstructural features which are
relevant for the mechanical properties. The number of straight
paths was found to follow an exponential distribution just like
the distribution of the force chain lengths in mechanically loaded
granular material~\cite{Peters_2005}. Additionally, the
quasi-linear structure of the straight paths seems to correspond
to the geometrical shape of force
chains~\cite{Drescher_1972,Cundall_1979} that are well known to
determine the load bearing capacity of granular matter. These
findings are encouraging for an attempt to establish a correlation
between defined microstructural features and the differences in
mechanical properties of destabilized colloids.

Further quantitative structural analyses and computational
modelling using the discrete element method are in progress to
study the force chains in ceramic green bodies and to correlate
them with the straight paths of their microstructures.

\begin{acknowledgments}
The authors would like to express their gratitude to Markus
H\"utter for providing the colloidal microstructure data from the
Brownian dynamics simulations.
\end{acknowledgments}

\newpage
\bibliography{Schenker_JEurCeramSoc_2008}

\begin{thebibliography}{24}
\expandafter\ifx\csname natexlab\endcsname\relax\def\natexlab#1{#1}\fi
\expandafter\ifx\csname bibnamefont\endcsname\relax
  \def\bibnamefont#1{#1}\fi
\expandafter\ifx\csname bibfnamefont\endcsname\relax
  \def\bibfnamefont#1{#1}\fi
\expandafter\ifx\csname citenamefont\endcsname\relax
  \def\citenamefont#1{#1}\fi
\expandafter\ifx\csname url\endcsname\relax
  \def\url#1{\texttt{#1}}\fi
\expandafter\ifx\csname urlprefix\endcsname\relax\def\urlprefix{URL }\fi
\providecommand{\bibinfo}[2]{#2}
\providecommand{\eprint}[2][]{\url{#2}}

\bibitem[{\citenamefont{Yun et~al.}(2007)\citenamefont{Yun, Santamarina, and
  Ruppel}}]{Yun_2007}
\bibinfo{author}{\bibfnamefont{T.~S.} \bibnamefont{Yun}},
  \bibinfo{author}{\bibfnamefont{J.~C.} \bibnamefont{Santamarina}},
  \bibnamefont{and} \bibinfo{author}{\bibfnamefont{C.}~\bibnamefont{Ruppel}},
  \bibinfo{journal}{J. Geophys. Res.} \textbf{\bibinfo{volume}{112}},
  \bibinfo{pages}{B04106} (\bibinfo{year}{2007}).

\bibitem[{\citenamefont{Feng and Dogan}(2000)}]{Feng_2000}
\bibinfo{author}{\bibfnamefont{J.-H.} \bibnamefont{Feng}} \bibnamefont{and}
  \bibinfo{author}{\bibfnamefont{F.}~\bibnamefont{Dogan}},
  \bibinfo{journal}{Mater. Sci. Eng. A} \textbf{\bibinfo{volume}{283}},
  \bibinfo{pages}{56} (\bibinfo{year}{2000}).

\bibitem[{\citenamefont{Adeyeye et~al.}(2002)\citenamefont{Adeyeye, Jain,
  Ghorab, and Jr}}]{Adeyeye_2002}
\bibinfo{author}{\bibfnamefont{M.~C.} \bibnamefont{Adeyeye}},
  \bibinfo{author}{\bibfnamefont{A.~C.} \bibnamefont{Jain}},
  \bibinfo{author}{\bibfnamefont{M.~K.~M.} \bibnamefont{Ghorab}},
  \bibnamefont{and} \bibinfo{author}{\bibfnamefont{W.~J.~R.} \bibnamefont{Jr}},
  \bibinfo{journal}{AAPS PharmSciTech} \textbf{\bibinfo{volume}{3}},
  \bibinfo{pages}{8} (\bibinfo{year}{2002}).

\bibitem[{\citenamefont{Marti et~al.}(2005)\citenamefont{Marti, H{\"{o}}fler,
  Fischer, and Windhab}}]{Marti_2005}
\bibinfo{author}{\bibfnamefont{I.}~\bibnamefont{Marti}},
  \bibinfo{author}{\bibfnamefont{O.}~\bibnamefont{H{\"{o}}fler}},
  \bibinfo{author}{\bibfnamefont{P.}~\bibnamefont{Fischer}}, \bibnamefont{and}
  \bibinfo{author}{\bibfnamefont{E.~J.} \bibnamefont{Windhab}},
  \bibinfo{journal}{Rheol. Acta.} \textbf{\bibinfo{volume}{44}},
  \bibinfo{pages}{502} (\bibinfo{year}{2005}).

\bibitem[{\citenamefont{Gauckler et~al.}(1999)\citenamefont{Gauckler, Graule,
  and Baader}}]{Gauckler_1999}
\bibinfo{author}{\bibfnamefont{L.~J.} \bibnamefont{Gauckler}},
  \bibinfo{author}{\bibfnamefont{T.}~\bibnamefont{Graule}}, \bibnamefont{and}
  \bibinfo{author}{\bibfnamefont{F.}~\bibnamefont{Baader}},
  \bibinfo{journal}{Mater. Chem. Phys.} \textbf{\bibinfo{volume}{61}},
  \bibinfo{pages}{78} (\bibinfo{year}{1999}).

\bibitem[{\citenamefont{Tervoort et~al.}(2004)\citenamefont{Tervoort, Tervoort,
  and Gauckler}}]{Tervoort_2004}
\bibinfo{author}{\bibfnamefont{E.}~\bibnamefont{Tervoort}},
  \bibinfo{author}{\bibfnamefont{T.~A.} \bibnamefont{Tervoort}},
  \bibnamefont{and} \bibinfo{author}{\bibfnamefont{L.~J.}
  \bibnamefont{Gauckler}}, \bibinfo{journal}{J. Am. Ceram. Soc.}
  \textbf{\bibinfo{volume}{87}}, \bibinfo{pages}{1530} (\bibinfo{year}{2004}).

\bibitem[{\citenamefont{Wyss et~al.}(2004)\citenamefont{Wyss, Tervoort, Meier,
  M{\"{u}}ller, and Gauckler}}]{Wyss_2004}
\bibinfo{author}{\bibfnamefont{H.~M.} \bibnamefont{Wyss}},
  \bibinfo{author}{\bibfnamefont{E.}~\bibnamefont{Tervoort}},
  \bibinfo{author}{\bibfnamefont{L.~P.} \bibnamefont{Meier}},
  \bibinfo{author}{\bibfnamefont{M.}~\bibnamefont{M{\"{u}}ller}},
  \bibnamefont{and} \bibinfo{author}{\bibfnamefont{L.~J.}
  \bibnamefont{Gauckler}}, \bibinfo{journal}{J. Colloid Interface Sci.}
  \textbf{\bibinfo{volume}{273}}, \bibinfo{pages}{455} (\bibinfo{year}{2004}).

\bibitem[{\citenamefont{Wyss et~al.}(2001)\citenamefont{Wyss, Romer, Scheffold,
  Schurtenberger, and Gauckler}}]{Wyss_2001}
\bibinfo{author}{\bibfnamefont{H.~M.} \bibnamefont{Wyss}},
  \bibinfo{author}{\bibfnamefont{S.}~\bibnamefont{Romer}},
  \bibinfo{author}{\bibfnamefont{F.}~\bibnamefont{Scheffold}},
  \bibinfo{author}{\bibfnamefont{P.}~\bibnamefont{Schurtenberger}},
  \bibnamefont{and} \bibinfo{author}{\bibfnamefont{L.~J.}
  \bibnamefont{Gauckler}}, \bibinfo{journal}{J. Colloid Interface Sci.}
  \textbf{\bibinfo{volume}{241}}, \bibinfo{pages}{89} (\bibinfo{year}{2001}).

\bibitem[{\citenamefont{Balzer et~al.}(1999)\citenamefont{Balzer, Hruschka, and
  Gauckler}}]{Balzer_1999}
\bibinfo{author}{\bibfnamefont{B.}~\bibnamefont{Balzer}},
  \bibinfo{author}{\bibfnamefont{M.~K.~M.} \bibnamefont{Hruschka}},
  \bibnamefont{and} \bibinfo{author}{\bibfnamefont{L.~J.}
  \bibnamefont{Gauckler}}, \bibinfo{journal}{J. Colloid Interface Sci.}
  \textbf{\bibinfo{volume}{216}}, \bibinfo{pages}{379} (\bibinfo{year}{1999}).

\bibitem[{\citenamefont{Wyss et~al.}(2005)\citenamefont{Wyss, Deliormanli,
  Tervoort, and Gauckler}}]{Wyss_1_2005}
\bibinfo{author}{\bibfnamefont{H.~M.} \bibnamefont{Wyss}},
  \bibinfo{author}{\bibfnamefont{A.~M.} \bibnamefont{Deliormanli}},
  \bibinfo{author}{\bibfnamefont{E.}~\bibnamefont{Tervoort}}, \bibnamefont{and}
  \bibinfo{author}{\bibfnamefont{L.~J.} \bibnamefont{Gauckler}},
  \bibinfo{journal}{AIChE J.} \textbf{\bibinfo{volume}{51}},
  \bibinfo{pages}{134} (\bibinfo{year}{2005}).

\bibitem[{\citenamefont{Krall and Weitz}(1998)}]{Krall_1998}
\bibinfo{author}{\bibfnamefont{A.~H.} \bibnamefont{Krall}} \bibnamefont{and}
  \bibinfo{author}{\bibfnamefont{D.~A.} \bibnamefont{Weitz}},
  \bibinfo{journal}{Phys. Rev. Lett.} \textbf{\bibinfo{volume}{80}},
  \bibinfo{pages}{778} (\bibinfo{year}{1998}).

\bibitem[{\citenamefont{Hesselbarth}(2000)}]{Hesselbarth_2000}
\bibinfo{author}{\bibfnamefont{D.}~\bibnamefont{Hesselbarth}},
  \emph{\bibinfo{title}{Quellf{\"{a}}hige Polymerbinder in
  Aluminiumoxid-Suspensionen}} (\bibinfo{publisher}{Ph.D. thesis no. 13404, ETH
  Zurich, Switzerland}, \bibinfo{year}{2000}).

\bibitem[{\citenamefont{H{\"{u}}tter}(2000)}]{Huetter_2000}
\bibinfo{author}{\bibfnamefont{M.}~\bibnamefont{H{\"{u}}tter}},
  \bibinfo{journal}{J. Colloid Interface Sci.} \textbf{\bibinfo{volume}{231}},
  \bibinfo{pages}{337} (\bibinfo{year}{2000}).

\bibitem[{\citenamefont{H{\"{u}}tter}(2003)}]{Huetter_2003}
\bibinfo{author}{\bibfnamefont{M.}~\bibnamefont{H{\"{u}}tter}},
  \bibinfo{journal}{Phys. Rev. E} \textbf{\bibinfo{volume}{68}},
  \bibinfo{pages}{031404} (\bibinfo{year}{2003}).

\bibitem[{\citenamefont{H{\"{u}}tter}(1999)}]{Huetter_1999}
\bibinfo{author}{\bibfnamefont{M.}~\bibnamefont{H{\"{u}}tter}},
  \emph{\bibinfo{title}{Brownian Dynamics Simulation of Stable and of
  Coagulating Colloids in Aqueous Suspension}} (\bibinfo{publisher}{Ph.D.
  thesis no. 13107, ETH Zurich, Switzerland}, \bibinfo{year}{1999}).

\bibitem[{\citenamefont{Wyss et~al.}(2002)\citenamefont{Wyss, H{\"{u}}tter,
  M{\"{u}}ller, Meier, and Gauckler}}]{Wyss_2002}
\bibinfo{author}{\bibfnamefont{H.~M.} \bibnamefont{Wyss}},
  \bibinfo{author}{\bibfnamefont{M.}~\bibnamefont{H{\"{u}}tter}},
  \bibinfo{author}{\bibfnamefont{M.}~\bibnamefont{M{\"{u}}ller}},
  \bibinfo{author}{\bibfnamefont{L.~P.} \bibnamefont{Meier}}, \bibnamefont{and}
  \bibinfo{author}{\bibfnamefont{L.~J.} \bibnamefont{Gauckler}},
  \bibinfo{journal}{J. Colloid Interface Sci.} \textbf{\bibinfo{volume}{248}},
  \bibinfo{pages}{340} (\bibinfo{year}{2002}).

\bibitem[{\citenamefont{Drescher and de~Josselin~de
  Jong}(1972)}]{Drescher_1972}
\bibinfo{author}{\bibfnamefont{A.}~\bibnamefont{Drescher}} \bibnamefont{and}
  \bibinfo{author}{\bibfnamefont{G.}~\bibnamefont{de~Josselin~de Jong}},
  \bibinfo{journal}{J. Mech. Phys. Solids} \textbf{\bibinfo{volume}{20}},
  \bibinfo{pages}{337} (\bibinfo{year}{1972}).

\bibitem[{\citenamefont{Cundall and Strack}(1979)}]{Cundall_1979}
\bibinfo{author}{\bibfnamefont{P.~A.} \bibnamefont{Cundall}} \bibnamefont{and}
  \bibinfo{author}{\bibfnamefont{O.~D.~L.} \bibnamefont{Strack}},
  \bibinfo{journal}{G{\'{e}}otechnique} \textbf{\bibinfo{volume}{29}},
  \bibinfo{pages}{47} (\bibinfo{year}{1979}).

\bibitem[{\citenamefont{Russel et~al.}(March 1989)\citenamefont{Russel,
  Saville, and Schowalter}}]{Russel_1989}
\bibinfo{author}{\bibfnamefont{W.~B.} \bibnamefont{Russel}},
  \bibinfo{author}{\bibfnamefont{D.~A.} \bibnamefont{Saville}},
  \bibnamefont{and} \bibinfo{author}{\bibfnamefont{W.~R.}
  \bibnamefont{Schowalter}}, \emph{\bibinfo{title}{Colloidal Dispersions}}
  (\bibinfo{publisher}{Cambridge University Press}, \bibinfo{year}{March
  1989}).

\bibitem[{\citenamefont{Clarke and J{\'{o}}nsson}(1993)}]{Clarke_1993}
\bibinfo{author}{\bibfnamefont{A.~S.} \bibnamefont{Clarke}} \bibnamefont{and}
  \bibinfo{author}{\bibfnamefont{H.}~\bibnamefont{J{\'{o}}nsson}},
  \bibinfo{journal}{Phys. Rev. E} \textbf{\bibinfo{volume}{47}},
  \bibinfo{pages}{3975} (\bibinfo{year}{1993}).

\bibitem[{\citenamefont{Wielopolski and Smith}(1986)}]{Wielopolski_1986}
\bibinfo{author}{\bibfnamefont{P.~A.} \bibnamefont{Wielopolski}}
  \bibnamefont{and} \bibinfo{author}{\bibfnamefont{E.~R.} \bibnamefont{Smith}},
  \bibinfo{journal}{J. Chem. Phys.} \textbf{\bibinfo{volume}{84}},
  \bibinfo{pages}{6940} (\bibinfo{year}{1986}).

\bibitem[{\citenamefont{Aste}(2005)}]{Aste_2005}
\bibinfo{author}{\bibfnamefont{T.}~\bibnamefont{Aste}}, \bibinfo{journal}{J.
  Phys.: Condens. Matter} \textbf{\bibinfo{volume}{17}}, \bibinfo{pages}{S2361}
  (\bibinfo{year}{2005}).

\bibitem[{\citenamefont{Aste}(2006)}]{Aste_2006}
\bibinfo{author}{\bibfnamefont{T.}~\bibnamefont{Aste}}, \bibinfo{journal}{Phys.
  Rev. Lett.} \textbf{\bibinfo{volume}{96}}, \bibinfo{pages}{018002}
  (\bibinfo{year}{2006}).

\bibitem[{\citenamefont{Peters et~al.}(2005)\citenamefont{Peters, Muthuswamy,
  Wibowo, and Tordesillas}}]{Peters_2005}
\bibinfo{author}{\bibfnamefont{J.~F.} \bibnamefont{Peters}},
  \bibinfo{author}{\bibfnamefont{M.}~\bibnamefont{Muthuswamy}},
  \bibinfo{author}{\bibfnamefont{J.}~\bibnamefont{Wibowo}}, \bibnamefont{and}
  \bibinfo{author}{\bibfnamefont{A.}~\bibnamefont{Tordesillas}},
  \bibinfo{journal}{Phys. Rev. E} \textbf{\bibinfo{volume}{72}},
  \bibinfo{pages}{041307} (\bibinfo{year}{2005}).

\end{thebibliography}

\end{document}